# Towards 100 Gbps Ethernet: Development of Ethernet / Physical Layer Aspects

Ömer Bulakci

*Abstract*— Physical layer features of Ethernet from the first realization towards the 100 Gb Ethernet (100 GbE) development have been considered. Comparisons of these features are made according to the standardized data rates. Feasible physical layer options are then discussed for high data rates.

## I. INTRODUCTION

ETHERNET is the most widely deployed Local Area Network (LAN) protocol and has been extended to Metropolitan Area Networks (MAN) and Wide Area Networks (WAN) [1]. The major advantages that characterize Ethernet can be stated as its cost efficiency, traditional tenfold bit rate increase (from 10 Mbps to 100 Gbps), simplicity, high transmission reliability and worldwide interoperability between vendors [2].

The first experimental Ethernet was developed during the early 1970s by XEROX Corporation in a coaxial cable network with a data rate about 3 Mbps [3]. The initial standardization process of Ethernet was started in 1979 by Digital Equipment Corporation (DEC), Intel and Xerox. In 1980, DIX Standard known as the "Thick Ethernet" was released. Thick Ethernet defined a bit rate of 10 Mbps and employed "thick" coaxial cable to connect the devices on the network and hence took its name from its transmission medium. The DIX Standard was revised by Institute of Electrical and Electronic Engineers (IEEE) in 1983 and the first IEEE standard, namely 802.3, was released. Table I summarizes the milestones in the 802.3 standard after the first release. In the table TP, MMF and SMF stand for Twisted Pair, Multi-Mode Fiber and Single-Mode Fiber, respectively. For 100 GbE Higher Speed Study Group (HSSG) was formed in 2006. HSSG determines the objectives for the new standard clause which is expected to come out in 2010 [4-6].

The paper is organized as follows. The second section of this paper presents various physical layer aspects of mentioned standards. Already implemented architectures for released clauses and ongoing architecture researches for 100 GbE are elaborated.

TABLE I

Milestones of 802.3 IEEE Standard

| Clause Name | Date of Release | Bit Rate | Physical Medium |
|---|---|---|---|
| 802.3a (Thin Ethernet) (Cheapernet) | 1985 | 10 Mbps | Single Thin Coaxial Cable |
| 802.3i | 1990 | 10 Mbps | TP Copper |
| 802.3j | 1993 | 10 Mbps | Two MMFs |
| 802.3u (Fast Ethernet) | 1995 | 100 Mbps | TP Copper Two Fibers (MMF,SMF) |
| 802.3z (Gigabit Ethernet) | 1998 | 1 Gbps | MMF, SMF |
| 802.3ab | 1999 | 1 Gbps | TP Copper |
| 802.3ae | 2002 | 10 Gbps | MMF,SMF |
| 802.3an | 2006 | 10 Gbps | TP Copper |
| 802.3ba | ~2010 | 100 Gbps | MMF,SMF, Copper |

## II. PHYSICAL LAYER ASPECTS

The first physical medium used for original 10 Mbps Ethernet is thick coaxial cable (10 mm) and is identified as 10BASE5 which is the shorthand for 10 Mbps BASEband transmission on a maximum supported cable segment length of 500 meters. To build a larger network, after each segment a repeater has to be used for regeneration of the transmit signal which is attenuated through the segment. However, thick and inflexible structure of the thick coaxial cable results in expensive and difficult installation of the network. Therefore, for the next clause thin coaxial cable with an identifier of 10BASE2 is used which is cheaper and more flexible. Nevertheless, due to the worse transmission characteristics it can support a maximum segment length of 185 meters. Although the installation of thin coaxial media is easier, management of Ethernet systems based on coaxial media is still difficult because of the problems of classical bus topology. A new era for Ethernet was started by means of the development of Twisted-Pair Ethernet, specifically 10BASE-T. 10BASE-T is based on star-wired topology and hence is easier to be managed. With 10BASE-T, Ethernet usage has widely and rapidly increased. The major drawback of Twisted-Pair Ethernet is the limited reach of a segment which is 100 meters. Longer reach systems are formed by fiber cabling identified as 10BASE-F which uses the light with a wavelength of 850 nm. A maximum segment length of 2000 meters can be supported by 10BASE-F using two MMFs with an optic core diameter of 62.5 microns and cladding diameter of 125 microns. All of these 10 Mbps schemes utilize Manchester Encoding as signaling. Manchester Encoding is a digital phase modulation encoding method where the clock of the signal can be recovered from the data signal. However, the decoding circuitry is complex and the modulation bandwidth is doubled compared to Non Return to Zero (NRZ) [7]. Fig. 1 shows an example of Manchester Encoded data stream.

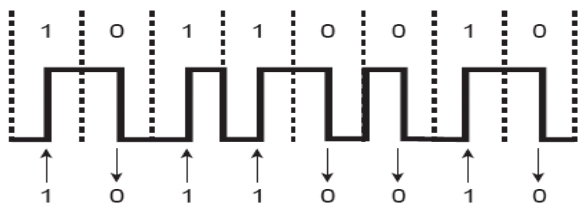

Fig.1    Manchester Encoded Baseband Transmission [3]

As the computer technology evolved the necessity of faster Ethernet increased. To meet the increased demand, the bit rate of Ethernet was increased by a factor of ten and 100 Mbps systems (Fast Ethernet) were developed. The physical media of Fast Ethernet still consist of Twisted-Pair (100BASE-T) and fiber optic cabling. To fulfill bandwidth requirements, a different category of Twisted-Pair cabling has to be used (Category 5, Unshielded Twisted-Pair ) , whereas fiber optic cabling structure utilizes the same MMF (62.5 /125) as before with an increased light wavelength of 1300 nm (100BASE-FX). The maximum supported segment lengths of these schemes remain the same as respective 10 Mbps schemes. The reach of the fiber optic cabling can be further increased up to 10 km by using SMF (100BASE-BX10). Both 100BASE-TX and 100BASE-FX use 4B/5B encoding for clock recovery where 4 bits words are assigned to 5 bits words in such a way that it guarantees transitions in each word. One extra bit is the overhead and can be used for error correction purposes. In addition to 4B/5B encoding, 100BASE-FX uses simple NRZ On-Off Keying (OOK) which is shown in Fig. 2 and 100BASE-TX uses MLT-3 encoding where it moves to one of the three states (-1,0,+1) for transmission of '1-bit' and remains at the initial state for transmission of '0-bit'.

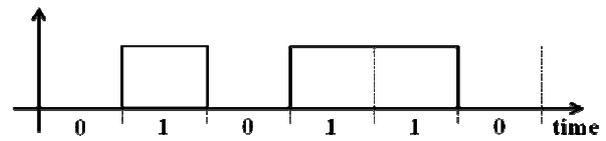

Fig.2    NRZ OOK Baseband Transmission

In the late '90s another tenfold increase was necessary to meet the drastic increase in the demand. Gigabit Ethernet was first developed on fiber-optic media for long reaches (1000BASE-SX, 1000BASE-LX), and shielded copper cable for short run connections (1000BASE-CX). 1000BASE-CX is shielded in order to prevent any interference and therefore differs from the classical UTP topology. 1000BASE-SX uses short wavelength laser technology which is cheaper, but has shorter reach and is only supported by MMF. 1000 BASE-LX uses long wavelength laser technology which is more expensive but has longer reach and is supported by both SMF and MMF. Longer reach of this type medium is due to utilization of attenuation dips in the fiber characteristics. The encoding used in the fiber channel is 8B/10B scheme which is quite similar to 4B/5B case; moreover, this encoding scheme has the property of DC balancing which helps preventing data dependent heating in the lasers [8]. The first release of Gigabit Ethernet did not introduce long-haul copper solution. Since the most widely deployed cabling system is Category 5 UTP copper cabling, the next release considered the long-haul copper solution and identified as 1000BASE-T with a maximum reach of 100 meters. 1000BASE-T has the advantage of cost efficiency because of the allowance of using existing copper cable plants. However, it has some inherent problems which make the design challenging, such as frequency dependent attenuation, crosstalk between wire pairs, echo and impedance mismatches. To cope with these problems, Forward Error Correction (FEC) coding is used to decrease the effects of high noise and crosstalk, and signal equalization at the receiver is used to compensate signal distortion caused by the channel. 1000BASE-T uses PAM-5 encoding where a transmit symbol can be at either of 5 states (-2, -1, 0, 1, 2). Four of the states are used for data transmission and one state is used for FEC. PAM-5 offers better bandwidth utilization than ordinary binary signaling, since each symbol carries two bits of information [9].

Being accepted as a tradition in resemblance to Moore's Law, next Ethernet technology showed another tenfold increase in bit rate and achieved the bit rate of 10 Gbps. With this standard, Ethernet extends its area from LAN into MAN and WAN [10]. Similar to Gigabit Ethernet, 10 GbE standard first considered fiber optic media. 10 GbE deploys both MMF and SMF. Mostly deployed MMF type (TIA-492AAAC: 50/125 microns) is standardized by 10GBASE-S and operates at 850 nm short wavelength with an effective modal bandwidth of 2000 MHz.km .It can achieve a reach of 300 meters. On the other hand, 10 GbE deploys standard SMF (IEC 60793-2) which operates at either 1310 nm wavelength or 1550 nm wavelength. The SMF design of 10 GbE had to face with various challenges, such as attenuation, chromatic dispersion and polarization mode dispersion (PMD) in order to be able to achieve its long reaches. PMD is not a major problem of the fibers which comply with new standard, but some of the old fiber infrastructures which are installed before '90s. The PMD issue will be important especially for very high data rates greater than 40 Gbps. For 1310 nm wavelength, the attenuation is the limiting factor for the reach. The attenuation is 0.35 dB/km, which is above the limit of optical amplification. Therefore, electrical regeneration should be applied after the maximum supported segment length of 10 km. Although the system with electrical regeneration is the expensive solution compared to optical amplification, 1310 nm wavelength is still preferred because of the price and availability of the lasers. For 1550 nm wavelength, however, the attenuation is 0.25 dB/km and hence below the optical amplification limit. Thus, optical amplification solutions like Erbium Doped Fiber Amplifier (EDFA) can be preferred as a cost effective scheme. As a consequence chromatic dispersion becomes the limiting factor for 1550 nm wavelength and limits its reach to 40 km. In order to counteract the chromatic dispersion either optical dispersion compensation or electrical regeneration can be used [11]. Fiber optic solution of 10 GbE utilizes 64B/66B encoding which has an acceptable overhead of 3.13 %. The advantages of this scheme are DC balancing, clock recovery capability, and robustness against malicious attacks. Contrary to previous encoding schemes, 64B/66B uses self-synchronous scrambler where scrambled 64-bits codewords are transmitted after the addition of 2-bits preamble. This preamble is used for error correction and for defining the run length [12]. The codewords can be transmitted by using conventional intensity modulation schemes like NRZ OOK.

Considering the cost performance and widely deployment of copper media, 10 GbE is also standardized for Twisted-Pair copper solution (10GBASE-T). The standardization process of TP copper solution had taken four years, since developers had to mitigate several challenges specific to copper media. First of all, in contrary to fiber optic media, to increase the data rate in copper media, the available bandwidth has to be exploited further by means of more complicated signaling and signal processing techniques. Moreover, the cabling impairments become deleterious for high data rates and some extra precautions should be considered compared to previous schemes. Table II shows a comparison between 1000BASE-T and 10GBASE-T solutions [13]. In the table FEXT stands for Far-End Crosstalk.

TABLE II
Comparison of Twisted-Pair Copper Solutions

| 1000BASE-T | 10GBASE-T |
|---|---|
| PAM-5 | PAM-16 |
| Echo-Cancelled Transmission | Echo-Cancelled Transmission |
| ~80 MHz used Bandwidth | >500 MHz used Bandwidth |
| FEXT Cancellation suggested | FEXT Cancellation required |
| FEC | Advanced FEC |

As Table II indicates, for 10 GbE instead of 2 bits/symbol transmission (PAM-5), the number of modulation levels has to be increased and consequently 3.125 bits/symbol (PAM-16) transmission is obtained. To achieve $10^{-12}$ Bit Error Ratio (BER) objective, Low Density Parity Check (LDPC) code is applied which is a powerful FEC coding scheme and can essentially improve echo and crosstalk cancellation. Accordingly, LDPC increases the receiver sensitivity which is vital for 10 Gbps transmission. 10GBASE-T needs to use pre-equalization at the transmitter rather than equalization at the receiver, since the high number of errors in the received signal does not allow proper usage of common Decision Feedback Equalizer (DFE) at the receiver. As a result, 10GBASE-T employs Tomlinson-Harashima Precoder (THP), which can be seen as a similar DFE used at the transmitter for the purpose of easing the recovery of transmitted signals. Another challenge of 10GBASE-T is the Alien Crosstalk which results from the interaction of adjacent cables in the transmission channel (between 4 pairs of TP copper cables). To mitigate Alien Crosstalk, Category 6 (Class E) and Category 7 (Class F) cabling structures can be used. Fig. 3 shows the stages used to mitigate cabling impairments. This figure demonstrates the impact of the impairments of copper cabling on 10 GbE. In the figure NEXT and ISI denote Near-End Crosstalk and Intersymbol Interference, respectively. By means of the advancements in 10GBASE-T, it can achieve a reach of 55 to 100 meters [13].

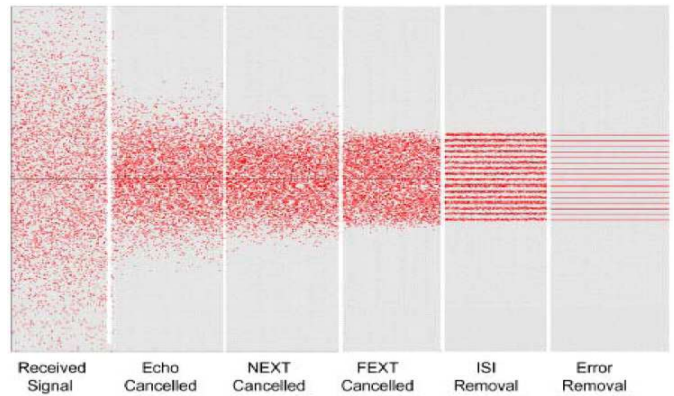

Fig. 3 Stages taken for correct detection of data symbols in 10GBASE-T [13]

The increasing traffic demand has raised the necessity of another tenfold bit rate increase in the speed of Ethernet. For this reason HSSG was formed and the target data rate of 100 Gbps was decided. Recently, Project Authorization Request (PAR) is approved for the further work, considering the Five Criteria (Broad Market Potential, Compatibility, Distinct Identity, Technical Feasibility, Economic Feasibility) [14]. HSSG defined the objectives of ongoing standardization [15]:
-At least a BER of $10^{-12}$
-A MAC data rate of 40 Gbps (higher at medium) over:
- at least 100 m of Optical Multimode 3 (OM3) MMF
- at least 10 m of copper cable assembly
- at least 1 m backplane

-A MAC data rate of 100 Gbps (higher at medium) over:
- at least 40 km of SMF
- at least 10 km of SMF
- at least 100 m of OM3 MMF
- at least 10 m of copper cable assembly

These objectives show that the concentration is mainly around fiber optic media. Nevertheless, copper media are of particular interest to have cost effective deployment. Current technology enables 100 Gbps transmission over a 5-m reach; however, for a reach of at least 10 m, high performance components are necessary with appropriate signaling schemes [16]. Apart from single 100 Gbps capable lanes, HSSG also considers different combinations of parallel aggregate lanes such as 10x10G, 5x20G, 4x25G and 2x50G. Fig. 4 shows two possible 100 GbE realizations [17]. The aggregation networks which consist of these aggregate lanes might be early generations of 100GbE in order to meet high demand. It seems to be that for data rates higher than 50 Gbps dispersion compensation would be a major issue [15],since effect of chromatic dispersion is proportional with (bit rate)$^2$ [17].

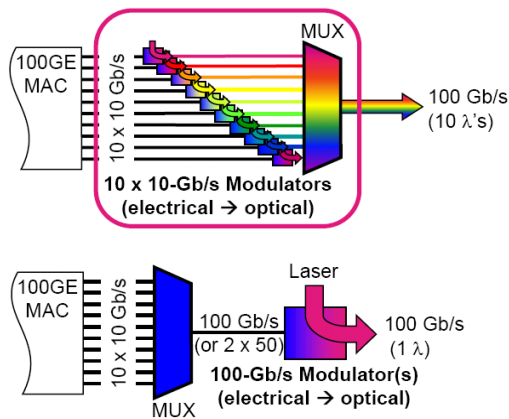

Fig. 4 100 GbE on Parallel Aggregate Lanes and Single Lane [17]

Current research considers different modulation formats for 100 GbE. Some of these formats are known from wireless technology and will be adapted to fiber optic media such as Phase Shift Keying (PSK) and Orthogonal Frequency Division Multiplexing (OFDM). To increase the spectral efficiency, the modulation format can be changed accordingly. PolMux (Polarization Multiplexing) seems to be an effective way to increase the spectral efficiency where phase modulation is applied to both of the optical fields ($E_x$, $E_y$) and hence 4 bits/symbols can be obtained by means of PolMux QPSK. However, PolMux and PSK schemes require more complicated hardware implementation capable of coherent detection. On the other hand, coherent detection increases the tolerance against chromatic dispersion and Polarization Mode Dispersion (PMD) [17]. In 2007, 10x111 Gbps (10 channels with 111Gbit/s each) transmission with PolMux-RZ (Return to Zero)-DQPSK (Differential Quadrature PSK) modulation was demonstrated with a reach of 2375 km [18]. PolMux OFDM seems to be an alternative scheme to PolMux QPSK. In OFDM data is distributed along the narrowband orthogonal subcarriers. A drawback of this scheme is the Peak-to-Average Power Ratio (PAPR) which is high for OFDM systems and decreases the resolution of DA/AD converters. Compared to PolMux QPSK modulation, recent experiments show that receiver sensitivity of OFDM is worse. In addition to these, OFDM systems are sensitive to frequency errors which may result in orthogonality loss. The advantage of OFDM is the high tolerance against chromatic dispersion and PMD, with simpler equalization. It should be noted that optical OFDM is a newer approach and is advancing with ongoing researches [19].

### III. CONCLUSION

It has been observed that from late '90s, every 2.5 years the traffic has increased with a factor of ten. Therefore, to meet this demand Ethernet technology has followed the tenfold increase in the data rate. As the data rate scales, fiber optic solutions become more attractive than the copper solutions due to higher reach, higher bandwidth and simpler signaling properties. In spite of that, copper solutions have to be considered in any case for cost effective deployment.